\pdfoutput=1
\newcommand*{\ATLASLATEXPATH}{latex/}
\documentclass[cernpreprint,texlive=2011,txfonts,UKenglish]{\ATLASLATEXPATH atlasdoc}

\usepackage{\ATLASLATEXPATH atlaspackage}
\usepackage{\ATLASLATEXPATH atlasbiblatex}

\usepackage{\ATLASLATEXPATH atlascontribute}

\usepackage{\ATLASLATEXPATH atlasphysics}

\graphicspath{{logos/}{figures/}}

\usepackage{subfig}

\def\missET {E_{\textrm{T}}^{\textrm{miss}}}
\def\missETnomu {E_{\textrm{T,no}\mu}^{\textrm{miss}}}
\def\missETvec {\textbf{E}_{\textrm{T}}^{\textrm{miss}}}
\def\missETnomuvec {\textbf{E}_{\textrm{T,no}\mu}^{\textrm{miss}}}

\DeclareGraphicsExtensions{.pdf,.png,.jpg,.jpeg,.eps}

\newcommand{\nb}{\ensuremath{\mathrm{nb}}}
\newcommand{\pb}{\ensuremath{\mathrm{pb}}}
\newcommand{\fb}{\ensuremath{\mathrm{fb}}}

\newcommand{\rom}[1]{\uppercase\expandafter{\romannumeral #1\relax}}

\AtlasTitle{Search for dark matter produced in association with a hadronically decaying vector boson in $pp$ collisions at $\sqrt{s}=13$~TeV with the ATLAS detector}

\author{The ATLAS Collaboration}

\AtlasRefCode{EXOT-2015-08}

\PreprintIdNumber{CERN-EP-2016-180}

\AtlasJournal{Phys.\ Lett.\ B.}
\AtlasJournalRef{Phys. Lett. B 763 (2016) 251}
\AtlasDOI{10.1016/j.physletb.2016.10.042}

\AtlasAbstract{%
A search is presented for dark matter produced in association with a hadronically decaying $W$ or $Z$ boson using 3.2 fb$^{-1}$  of $pp$ collisions at $\sqrt{s}=13$~TeV recorded by the ATLAS detector at the Large Hadron Collider. Events with a hadronic jet compatible
with a $W$ or $Z$ boson and with large missing transverse momentum are analysed. The data are
consistent with the Standard Model predictions and are interpreted in terms of both an effective field theory and a simplified model containing dark matter.
}

\hypersetup{pdftitle={ATLAS document},pdfauthor={The ATLAS Collaboration}}

\begin{document}

\maketitle

Dark matter is the dominant component of matter in the universe, but its particle nature remains a mystery.  Searches for a weakly interacting massive particle (WIMP), denoted by $\chi$, and for interactions between $\chi$ and Standard Model (SM) particles are a central component of the current set of dark-matter experiments.

\begin{figure}[thbp]
\begin{center}
\subfloat[]{
    \includegraphics[width=0.45\linewidth]{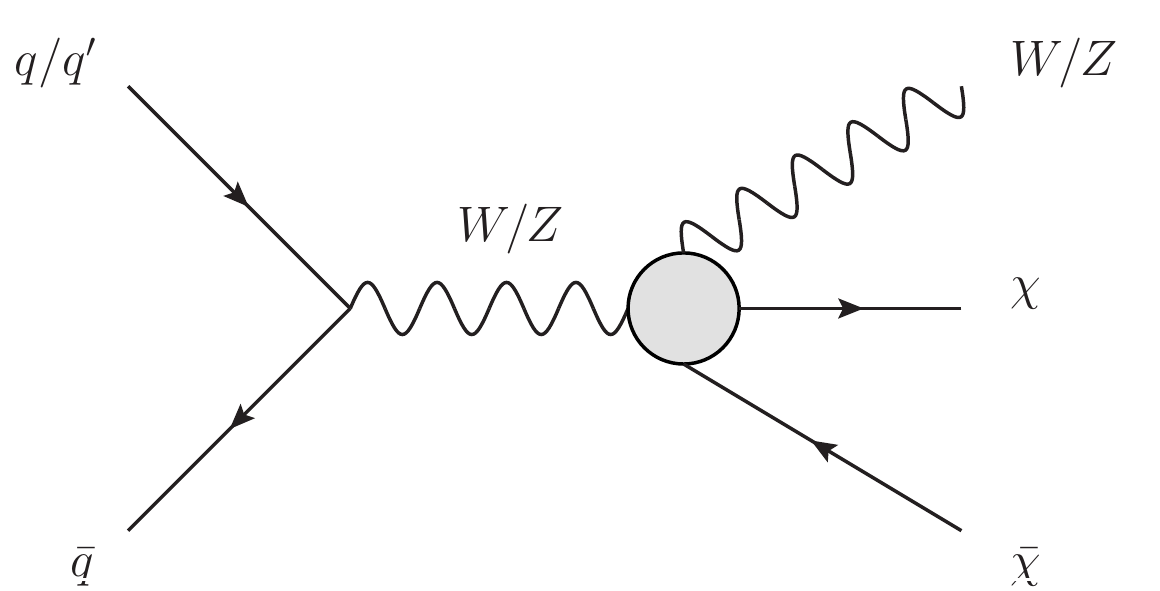}
}
\subfloat[]{
    \includegraphics[width=0.4\linewidth]{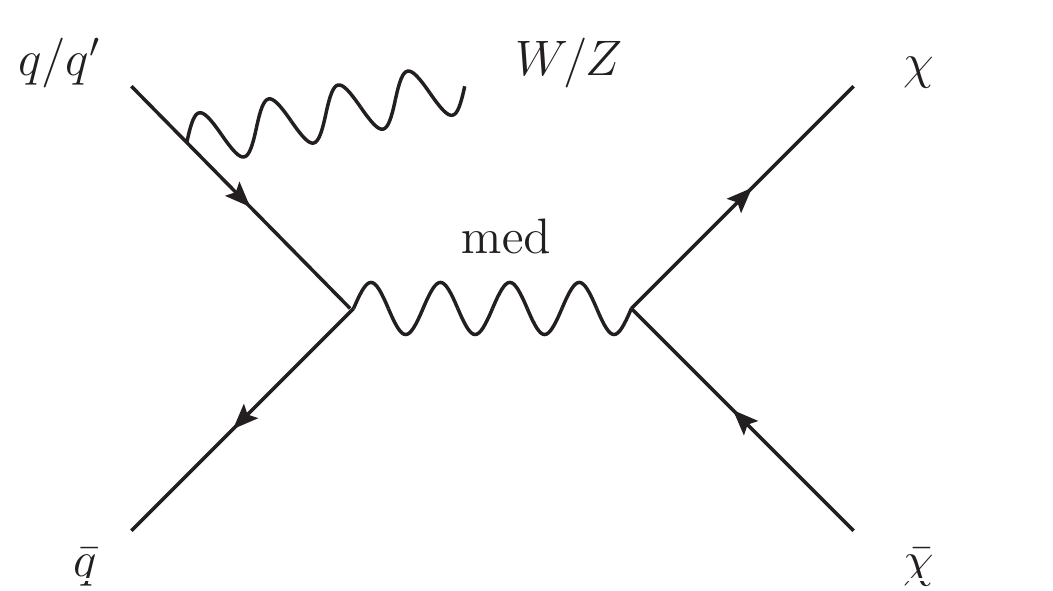}
}
\caption{ \label{fig:monow} Pair production of WIMPs ($\chi\bar{\chi}$) in proton--proton
  collisions at the LHC in association with a vector boson ($V$, meaning $W$ or $Z$) via two hypothetical processes: (a) production via an effective $VV\chi\chi$ interaction or (b) via a simplified model which includes an $s$-channel mediator.}
\end{center}
\end{figure}

At particle colliders, dark-matter particles may be produced in pairs via some unknown intermediate state.  While in many models direct detection experiments have the greatest sensitivity for dark-matter masses  $m_\chi$ between 10 and 100 GeV, searches for dark matter at particle colliders are most powerful for lower masses~\cite{cdfjmet,Khachatryan:2014rra,Aaboud:2016tnv}.  The final-state WIMPs are not directly detectable, but their presence can be inferred from the recoil against a visible particle~\cite{cdfjmet}. Two example processes are shown in Figure~\ref{fig:monow}.

The Tevatron and LHC collaborations have reported limits on the cross section of $p\bar{p}\rightarrow  \chi\bar{\chi}+X$ and 
$pp\rightarrow  \chi\bar{\chi}+X$, respectively, where $X$ is a hadronic jet~\cite{cdfjmet,Khachatryan:2014rra,Aaboud:2016tnv}, a photon ($\gamma$)~\cite{Aad:2012fw,Khachatryan:2014rwa}, a $W/Z$ boson~\cite{Aad:2014vka,Aad:2013oja}, or a Higgs boson~\cite{Aad:2015yga,monoHbb}.  In many cases, results are reported in terms of limits on the parameters of an effective field
theory (EFT) formulated as a four-point contact interaction~\cite{Cao:2009uw,Beltran:2010ww,Goodman:2010yf,Bai:2010hh,Goodman:2010ku,Rajaraman:2011wf,Fox:2011pm,Cheung:2012gi,Cotta:2012nj} between quarks and WIMPs. For such models, the strongest limits come from data in which the recoiling object is a jet. In other models, however, the interaction is between dark matter and vector bosons~\cite{Carpenter:2012rg}, such that the primary discovery mode would be in final states such as those analysed here, where the recoiling object is a $W$ or $Z$ boson.

In this letter,  a search is reported for  the production of a $W$ or $Z$
boson decaying hadronically (to $q\bar{q}'$ or $q\bar{q}$, respectively)
and reconstructed as a single massive jet in
association with large missing transverse momentum from the undetected
$\chi\bar{\chi}$ particles in data collected by the ATLAS detector from $pp$ collisions with centre-of-mass energy $\sqrt{s}=13$ TeV.  This search is sensitive to WIMP pair production,
as well as to other dark-matter-related models which predict invisible Higgs boson decays ($WH$ or $ZH$ production with
$H\rightarrow\chi\bar{\chi}$).

The ATLAS detector~\cite{atlas} at the LHC covers the
pseudorapidity\footnote{ATLAS uses a right-handed coordinate system
  with its origin at the nominal interaction point (IP) in the centre
  of the detector and the $z$-axis along the beam pipe. The $x$-axis
  points from the IP to the centre of the LHC ring, and the $y$-axis
  points upward. Polar coordinates ($r, \phi$) are used in the
  transverse ($x,y$) plane, $\phi$ being the azimuthal angle around
  the beam pipe. The pseudorapidity is defined in terms of the polar
  angle $\theta$ as $\eta = -\ln \tan(\theta/2)$.} range $|\eta| <
4.9$ and the full azimuthal angle $\phi$. It consists of an inner tracking detector
surrounded by a superconducting solenoid, electromagnetic and
hadronic calorimeters, and an external muon spectrometer incorporating
large superconducting toroidal magnets. A two-level trigger system
is used to select interesting events to be recorded for subsequent
offline analysis. Only data for which beams were stable and all subsystems described above
were operational are used. Applying these requirements to $pp$ collision
data, recorded during the 2015 LHC run, results in a data
sample with a time-integrated luminosity of $3.2$~fb$^{-1}$. The systematic uncertainty  of 2.1\% in the luminosity is derived following the same methodology as that detailed in Ref.~\cite{lumi}.

Three non-exclusive categories of jet candidates are built, each using the anti-$k_{\perp}$ clustering algorithm~\cite{Cacciari:2008gp}. Two categories use clusters of energy deposits in calorimeter cells seeded by those with
energies significantly above the measured noise and calibrated at the
hadronic energy scale~\cite{topocal}. They are distinguished by their radius parameters; jets with radius parameter of 1.0 (0.4) are referred to as  {\it large-R jets (narrow jets)}. Large and narrow jets can share a fraction of their energy deposits. A third type of jet candidate is reconstructed from inner-detector tracks using the anti-$k_{\perp}$ algorithm with $R=0.2$, referred to as {\it track jets}. Large-$R$ jets are trimmed~\cite{Krohn:2009th} to remove energy deposited by pile-up jets, the underlying event, and soft radiation. In this process, the constituents of large-$R$ jets are reclustered using the $k_{\perp}$ algorithm~\cite{PhysRevD.48.3160,CATANI1993187} with a distance parameter of $0.2$, and subjets
with transverse momentum $p_{\textrm{T}}$ less than 5\% of the large-$R$ jet $p_\textrm{T}$ are removed.  Large-$R$ jets are required to satisfy $p_{\textrm{T}}> 200$ GeV and $|\eta|<2.0$. These large-$R$ jets are intended to capture the
hadronic products of both quarks from the decay of a $W$ or $Z$ boson, while the narrow jets and track jets are helpful in background suppression. The internal structure of the large-$R$ jet is characterized in terms of two quantities:  $D_2$~\cite{d2variable,c2variable}, which identifies jets with two distinct concentrations of energy~\cite{Aad:2013gja,ATLAS-CONF-2015-035}, 
and $m_{\textrm{jet}}$, which is the calculated invariant mass of the jet. Narrow jets are required to satisfy $p_{\textrm{T}}> 20$ GeV for $|\eta|<2.5$ or  $p_{\textrm{T}}> 30$ GeV for $2.5<|\eta|<4.5$. Track jets are required to satisfy $p_{\textrm{T}}> 10$ GeV and $|\eta|<2.5$.   For both the large-$R$ and narrow jets, jet momenta are
calculated by performing a four-vector sum over these component
clusters, treating each topological cluster~\cite{topocal} as an $(E, \vec{p}\,)$ four-vector with zero
mass, and are calibrated to the hadronic scale. For narrow jets, the direction of $\vec{p}$ is given by the line joining the
reconstructed vertex with the barycentre of the energy cluster. The missing transverse momentum $\missETvec$ is calculated as the negative of the vector sum of the transverse momenta of reconstructed jets, leptons, and those tracks which are associated with the reconstructed vertex but not with any jet or lepton. A closely related quantity, $\missETnomuvec$, is calculated in the same way but excluding reconstructed muons.  A third variant,  $\textbf{p}_{\textrm{T}}^{\textrm{miss}}$, is the missing transverse momentum  measured using inner detector tracks. The magnitudes of the three missing-tranverse-momentum variants are denoted by $\missET$, $\missETnomu$, and $p_{\textrm{T}}^{\textrm{miss}}$, respectively. 
Electrons,
muons, jets, and $\missETvec$ are reconstructed as described in
Refs.~\cite{eledef,Aad:2016jkr,topocal,metdef}, respectively.

Candidate signal events are selected by an inclusive $\missET$
trigger that is more than 99\% efficient for events with
$\missET>200$~GeV. Events triggered by  detector noise and
non-collision backgrounds are rejected as described in Ref.~\cite{atlasjet}.  In addition, events are required to satisfy the requirements of $\missET>250$~GeV, no reconstructed electrons or muons, and at least one large-$R$ jet with $p_{\textrm T}>200$~GeV, $|\eta|<2.0$, $m_{\textrm{jet}}$ and $D_2$ consistent with a $W$ or $Z$ boson decay as in Ref.~\cite{Aaboud:2016okv}.   To further suppress backgrounds from multijet and $t\bar{t}$ production, events are required to satisfy $p_{\textrm{T}}^{\textrm{miss}}>30$~GeV, a minimum azimuthal angular distance, $\Delta\phi$, of 0.6 between the $\missETvec$ and the nearest narrow jet, and $\Delta\phi(\missETvec,\textbf{p}_{\textrm{T}}^{\textrm{miss}})<\pi/2$.     Within a fiducial volume defined at parton level by similar selection requirements (except those on $D_2$ and $p_{\textrm{T}}^{\textrm{miss}}$), the reconstruction efficiency for the signal models described above varies from 38\% to 49\%.

\begin{figure}
\begin{center}
\subfloat[]{
    \includegraphics[width=0.48\linewidth]{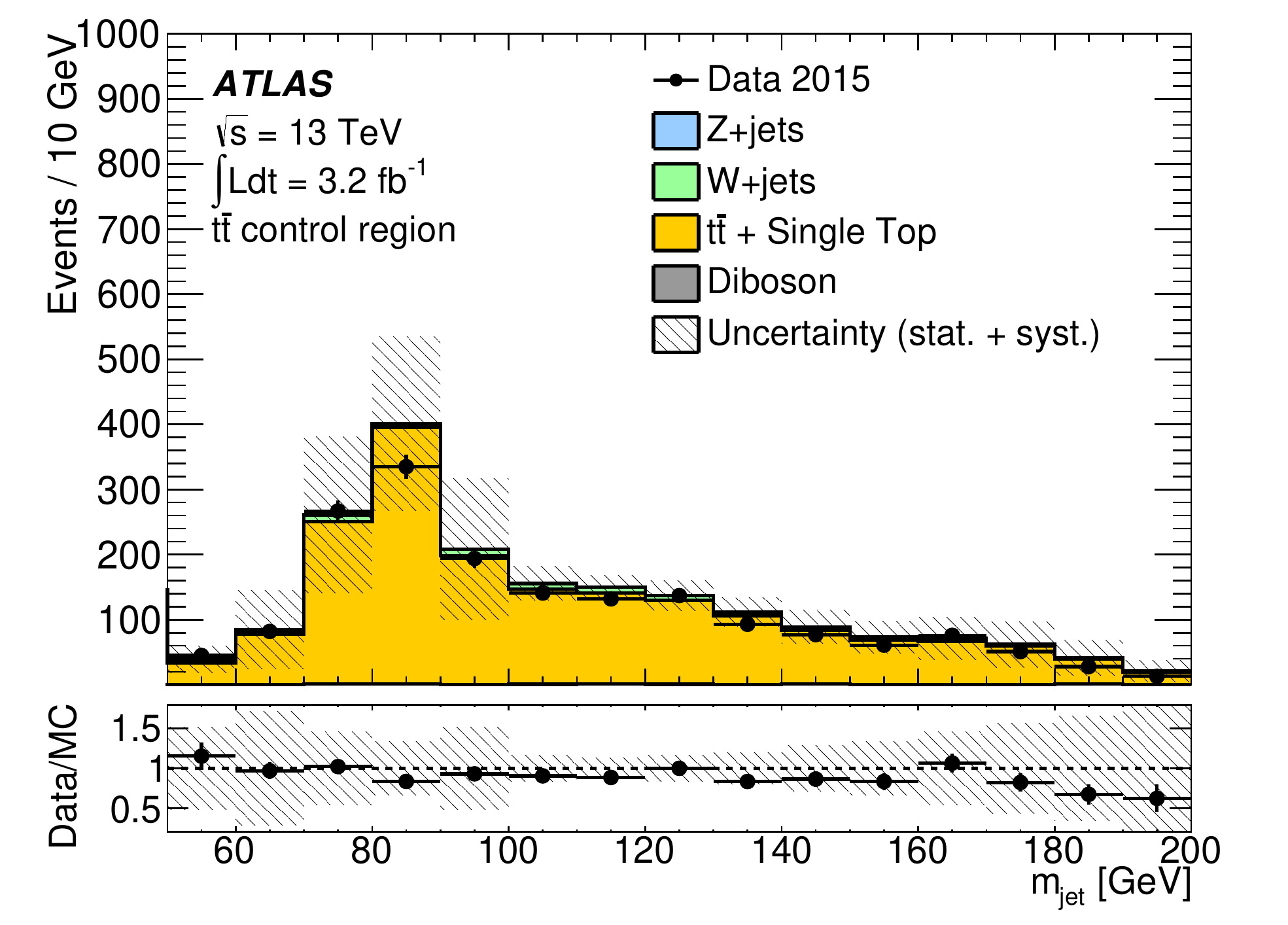}
}
\subfloat[]{
    \includegraphics[width=0.48\linewidth]{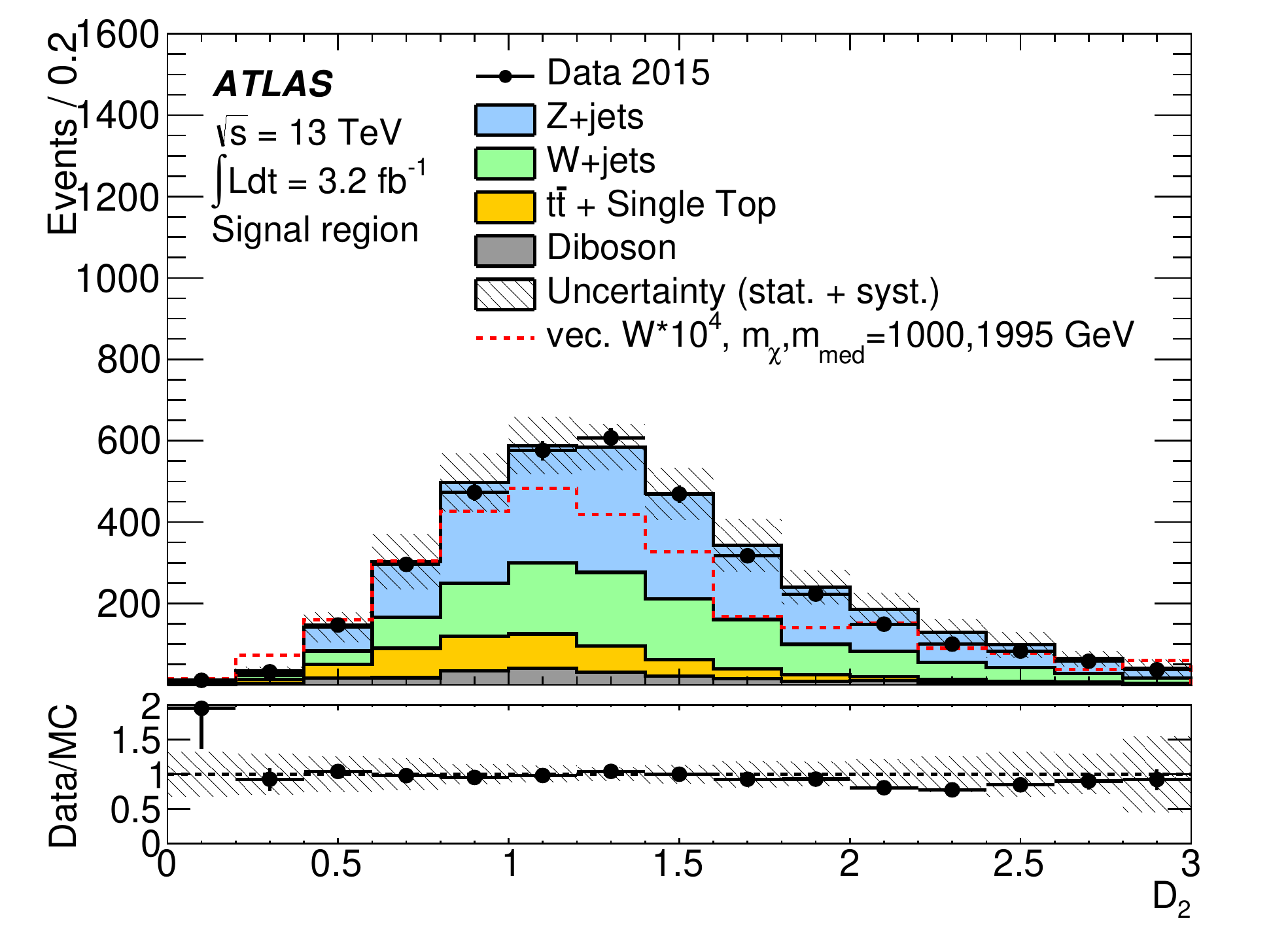}
}
\caption{ \label{fig:control} Pane (a) Distribution of $m_{\textrm{jet}}$ in the data and for the predicted background in the top-quark control region. Pane (b) Distribution of jet substructure variable $D_2$ in the data and for the predicted background in events satisfying all signal region requirements other than those on $D_2$. Also shown is the distribution for the simplified model with a vector-boson mediator, scaled by a factor of $10^4$ for given values of  $m_{\chi}$ and $m_{\textrm{med}}$, the mediator mass.
}
\end{center}
\end{figure}

The dominant source of background events is $Z\rightarrow \nu\bar{\nu}$
production in association with jets. A secondary
contribution comes from the production of jets in association with a  leptonically decaying
$W$ or $Z$ boson 
in which the charged leptons are not identified or the $\tau$ leptons decay hadronically.  The third major background contribution comes from top-quark pair production.  The kinematic distributions of these three largest backgrounds are estimated using simulated event samples but the  normalization is determined using control regions where the dark-matter signal is expected to be negligible. Each control region requires $\missET>200$~GeV and $p_{\textrm{T}}^{\textrm{miss}}>30$~GeV as well as one large-$R$ jet satisfying the substructure requirement on $D_2$ as applied in the signal region. The $Z$ boson control region requires exactly two muons with dimuon invariant mass $66 < m_{\mu\mu} < 116$~GeV. The $W$ boson (top quark) control region requires exactly one muon, and zero (at least one) $b$-tagged track jet not associated with the large-$R$ jet.    Validation of the  reconstruction of hadronic $W$ boson decays with
large-$R$ jets is performed in the top-quark control region, as shown in Figure~\ref{fig:control}, which also presents the distribution of the $D_2$ substructure variable.  Other sources of background are diboson production and single-top-quark production. The contribution to the signal region from multijet production is negligble.

Samples of simulated $W$+jets and $Z$+jets events are generated using {\sc Sherpa 2.1.1}~\cite{Gleisberg:2008ta}. Matrix elements are calculated for up to two partons at next-to-leading order (NLO) and four partons at leading order (LO) using the Comix~\cite{Gleisberg:2008fv} and OpenLoops~\cite{Cascioli:2011va} matrix element generators and merged with the {\sc Sherpa} parton shower~\cite{Schumann:2007mg} using the ME+PS@NLO prescription~\cite{Hoeche:2012yf}. The CT10~\cite{Lai:2010vv} PDF set is used in conjunction with dedicated parton shower tuning developed by the {\sc Sherpa} authors. The $W/Z$ production rates are normalized to a next-to-next-to-leading order (NNLO) calculation~\cite{PhysRevD.74.114017}.
The production of $t\bar{t}$ and single-top processes, including $s$-channel, $t$-channel and $Wt$ production is modelled with the {\sc Powheg-Box} v2 generator~\cite{Nason:2004rx,Frixione:2007vw,Alioli:2010xd} interfaced to {\sc Pythia6.428}~\cite{Sjostrand:2006za}.  In these generators the CT10 and CTEQ6L1~\cite{Pumplin:2002vw} PDF sets are used, respectively. Top-quark pair production is normalized to NNLO with next-to-next-to-leading-logarithm corrections~\cite{Czakon:2013goa} in QCD while single-top processes are normalized at NLO~\cite{Alioli:2009je,Re:2010bp} in QCD.
  The diboson ($WW,WZ,ZZ$) processes are simulated using {\sc Sherpa 2.1.1} with the CT10 PDF and normalized at NLO~\cite{Campbell:2010ff,Campbell:2011bn} in QCD.  The multijet process is described using samples simulated with {\sc Pythia8.186}~\cite{Sjostrand:2007gs} and the NNPDF2.3LO~\cite{Ball:2012cx} PDF at leading order in QCD; these multijet samples were used to develop the background estimation strategy but not for the final background prediction.

Samples of simulated $W\chi\bar{\chi}$ and $Z\chi\bar{\chi}$ events are generated using {\sc MadGraph5\_aMC@NLO}~\cite{Alwall:2014hca}, and the underlying event and parton showering are simulated with {\sc Pythia8.186}~\cite{Sjostrand:2007gs}.  Two theoretical models are used as benchmarks: a seven-dimensional $VV\chi\chi$ EFT~\cite{Carpenter:2012rg} model ($V$ meaning $W$ or $Z$) and a vector-mediated simplified model~\cite{dmf}. 
The strength of the EFT interaction is controlled by a mass scale, $M_\star$, and 
the strength of the simplified model interaction is controlled by the product of the couplings of the mediator 
to the SM and the dark matter (DM) particles, $g_{\rm SM}g_{\rm DM}$.
The EFT model samples were generated with $M_\star=3000$~GeV, and the 
simplified model samples were generated with couplings $g_{\rm SM}=0.25$ and $g_{\rm DM}=1$.
The samples were generated  as a function of dark-matter particle mass $m_{\chi}$ for the EFT model and 
in a grid of mediator mass $m_{\mathrm{med}}$  and $m_{\chi}$ for the simplified model.

Major sources of systematic uncertainty are uncertainties in the modelling of large-$R$ jet observables, which have a 5--13\% impact on the expected background and signal yields, and the energy scale of the narrow jets, which contribute a 1--5\% uncertainty to the expected yields. Other sources of uncertainty include theoretical uncertainties in the simulated event samples used to model the background processes (1--10\%), parton distribution functions (10--15\%), and lepton reconstruction and identification efficiencies (up to 2\%).  

A profile-likelihood fit~\cite{Cowan:2010js} 
to the $\missET$ ($\missETnomu$) distribution in the signal region (control regions) is used to constrain the 
$W$ boson, $Z$ boson, 
and $t\bar{t}$ backgrounds and extract the signal strength, $\mu$, for each model as an overall normalization factor for the signal prediction.
Besides the signal strength, three overall normalization factors for the $W$ boson, $Z$ boson, 
and $t\bar{t}$ backgrounds are parameters in the fit.
The diboson and single-top backgrounds are 
estimated  from simulation, and the multijet background is negligible.
The likelihood function is defined as the product of 
Poisson distributions over all bins in $\missET$ and $\missETnomu$,
and the likelihood 
is simultaneously maximized over the signal and control regions.

Variations of the expected signal and background to allow for their systematic uncertainties are described with nuisance parameters
constrained by Gaussian
probability distribution functions, and 
correlations across signal and background processes and regions are taken into account.

\begin{figure}
\centering
\subfloat[]{
    \includegraphics[width=0.48\textwidth]{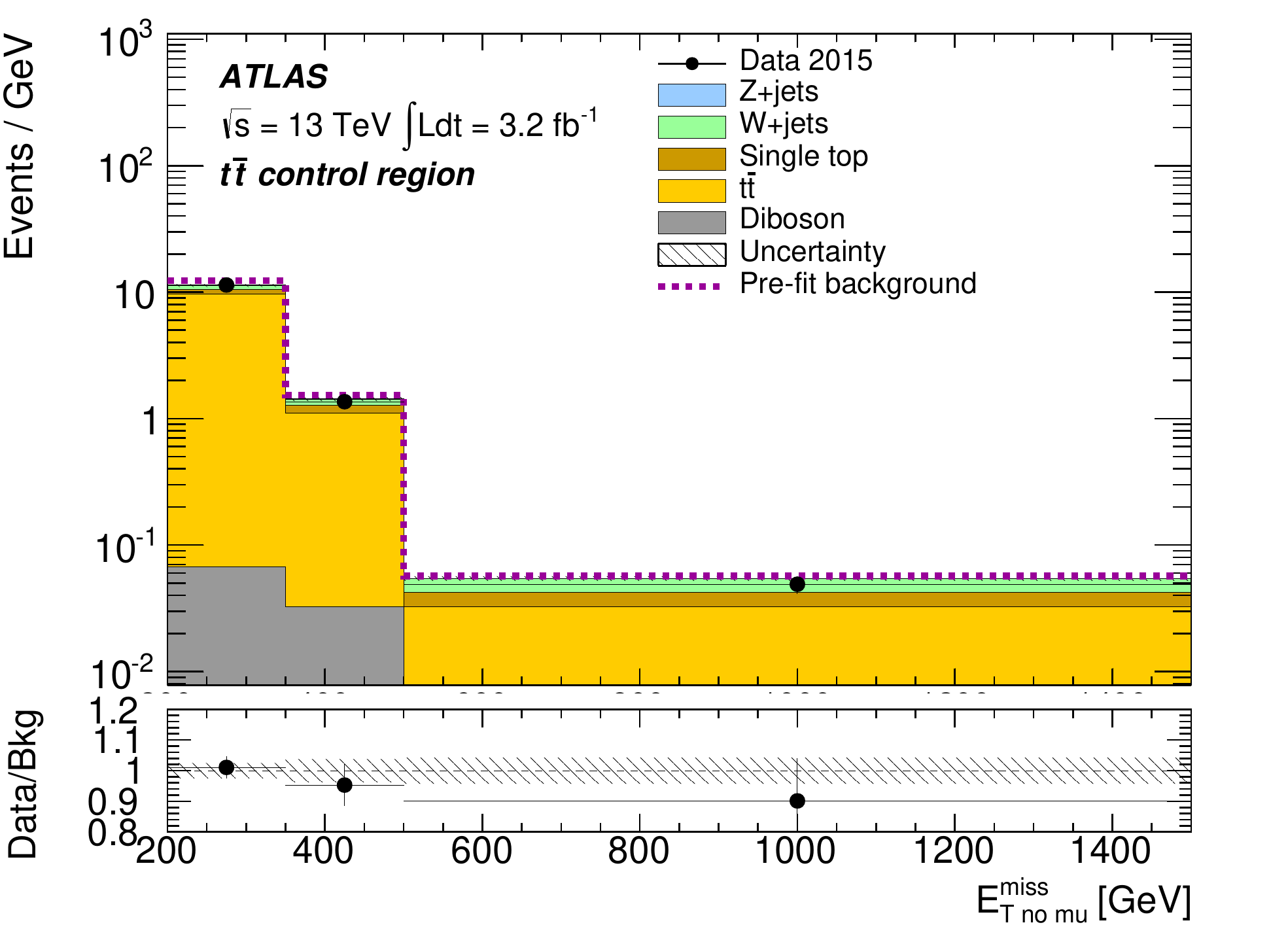}
}
\subfloat[]{
	\includegraphics[width=0.48\textwidth]{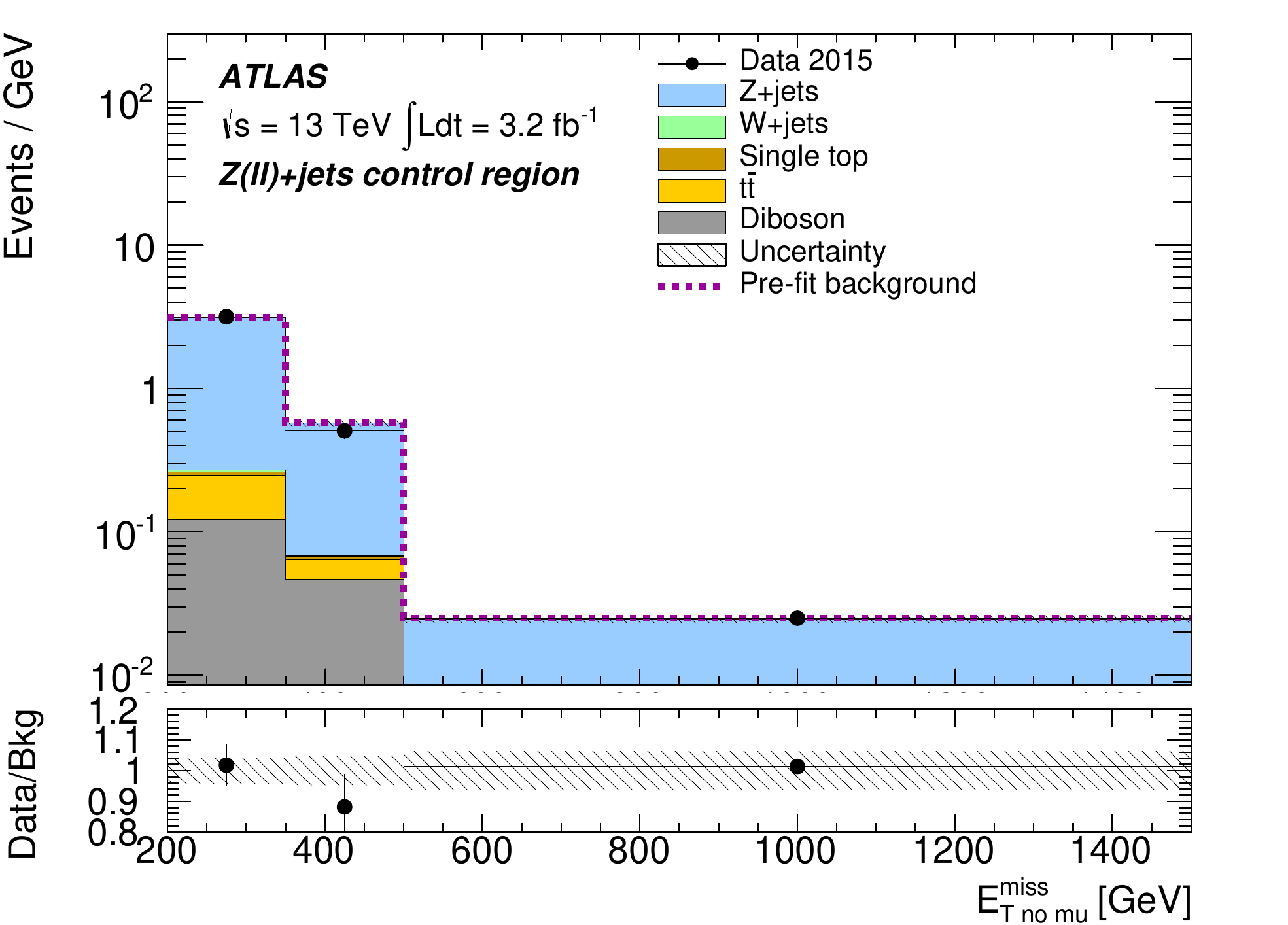}
}\\
\subfloat[]{
	\includegraphics[width=0.48\textwidth]{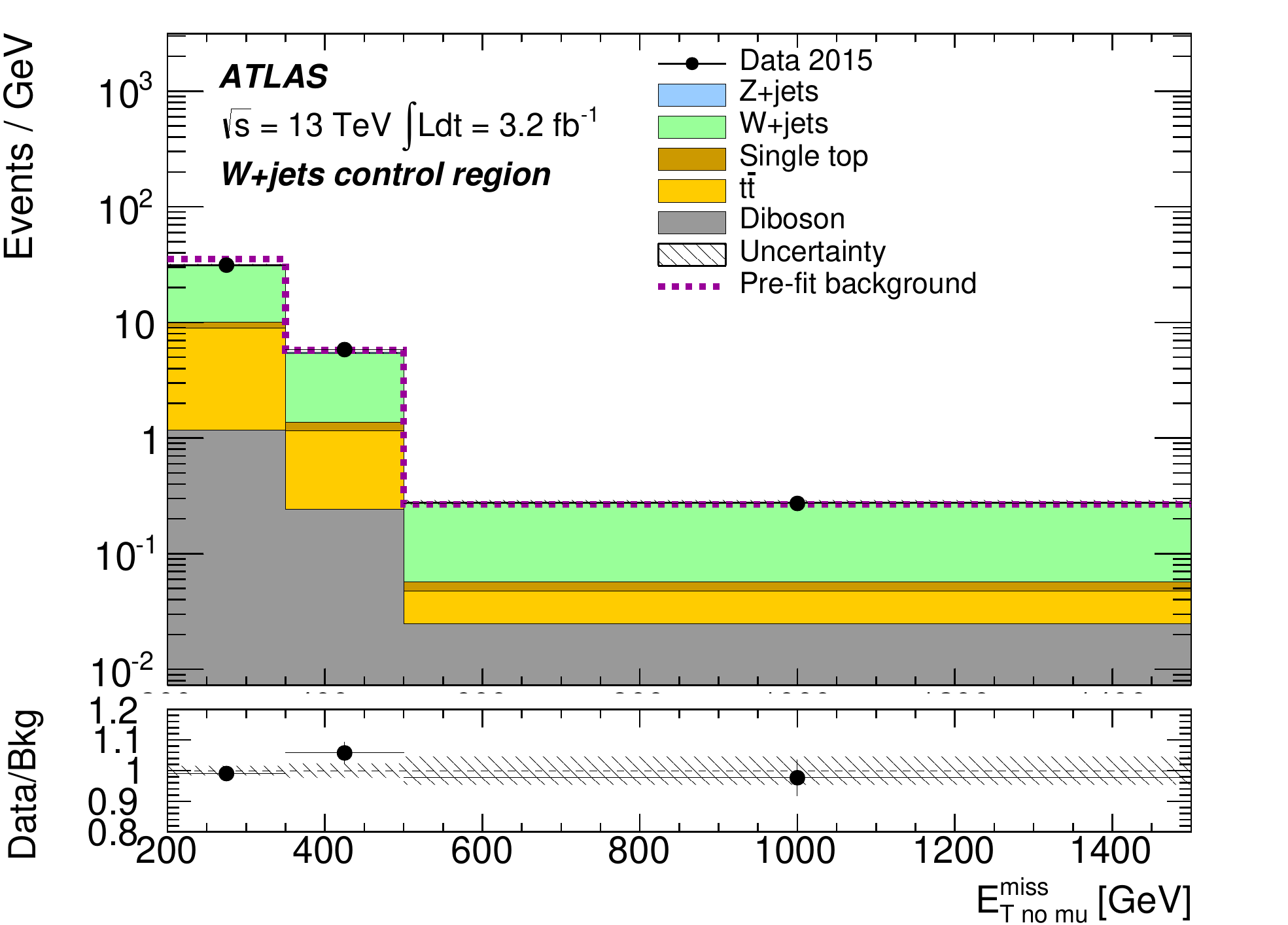}
	}
\caption{
\label{fig:postFit}
The $\missETnomu$  distribution of the events in the control regions after the profile-likelihood fit to the data under the background-only hypothesis.
Pane (a)  shows the $t\bar{t}$ control region, pane
(b) shows the $Z+$jets control region, and pane (c) shows the $W+$jets control region. The total background prediction before the fit is shown as a dashed line.
The inset at the bottom of each plot shows the ratio of the data to the total post-fit background. 
The hatched bands represent the total uncertainty in the background.
}
\end{figure}

\begin{figure}[h]
\centering
\includegraphics[width=0.48\textwidth]{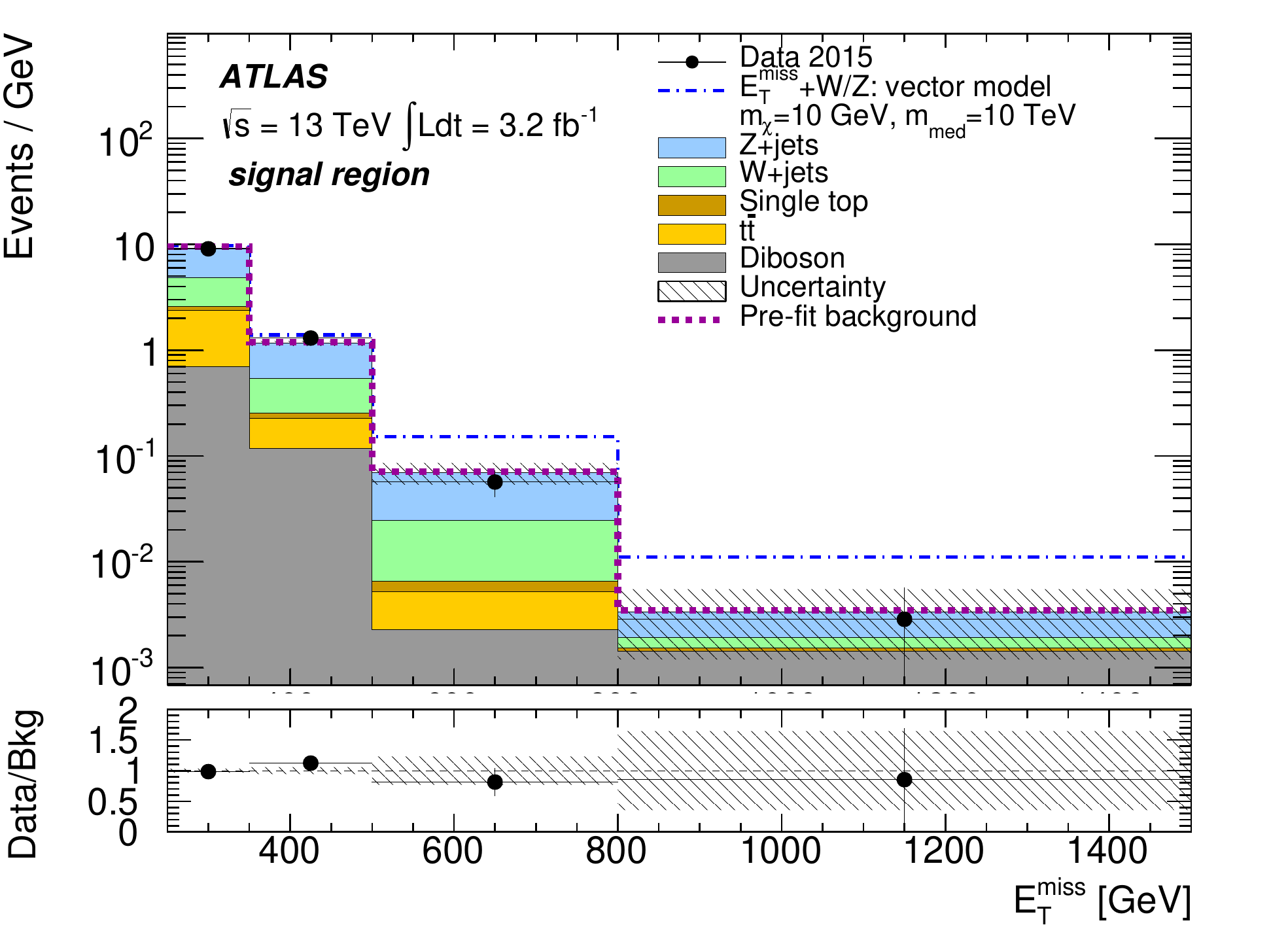}
\caption{
\label{fig:postFit2}
The $\missET$  distribution of the events in the signal region
after the profile-likelihood fit to the data under the background-only hypothesis. The inset shows the ratio of the data to the total background.  Also shown is the $\missET$ distribution for the simplified model with a vector-boson mediator, scaled by a factor of $10^4$ for  $m_{\chi}=10$~GeV and $m_{\textrm{med}}=10$~TeV.
The total background before the fit is shown as a dashed line.
The hatched bands represent the total uncertainty in the background.
}
\end{figure}

A background-only ($\mu=0$) fit, shows no 
deviation from SM predictions, and 
Figures~\ref{fig:postFit} and~\ref{fig:postFit2} show kinematic distributions 
after the profile-likelihood fit. The floating background-normalization parameters are consistent with unity within one standard deviation.
Tables~\ref{tab:event_yields} and~\ref{tab:normalization_factors} show the 
expected event yields after applying the signal selection and the background 
normalization scale factors, respectively. 
The values in these tables are estimated for the background-only hypothesis.

\begin{figure}[ht!]
\begin{center}
\subfloat[]{
\includegraphics[width=0.48\textwidth]{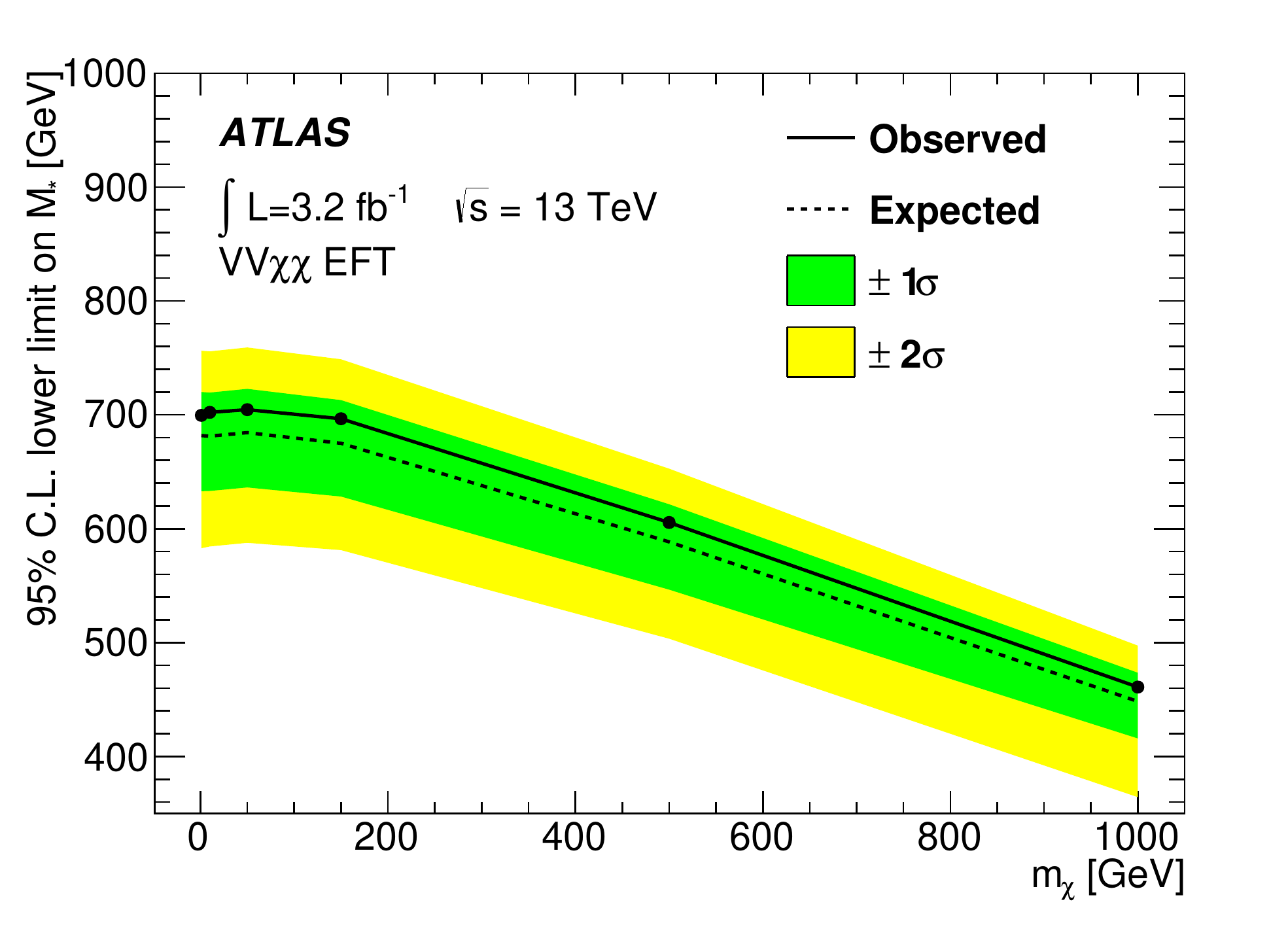} 
}
\subfloat[]{
\includegraphics[width=0.48\textwidth]{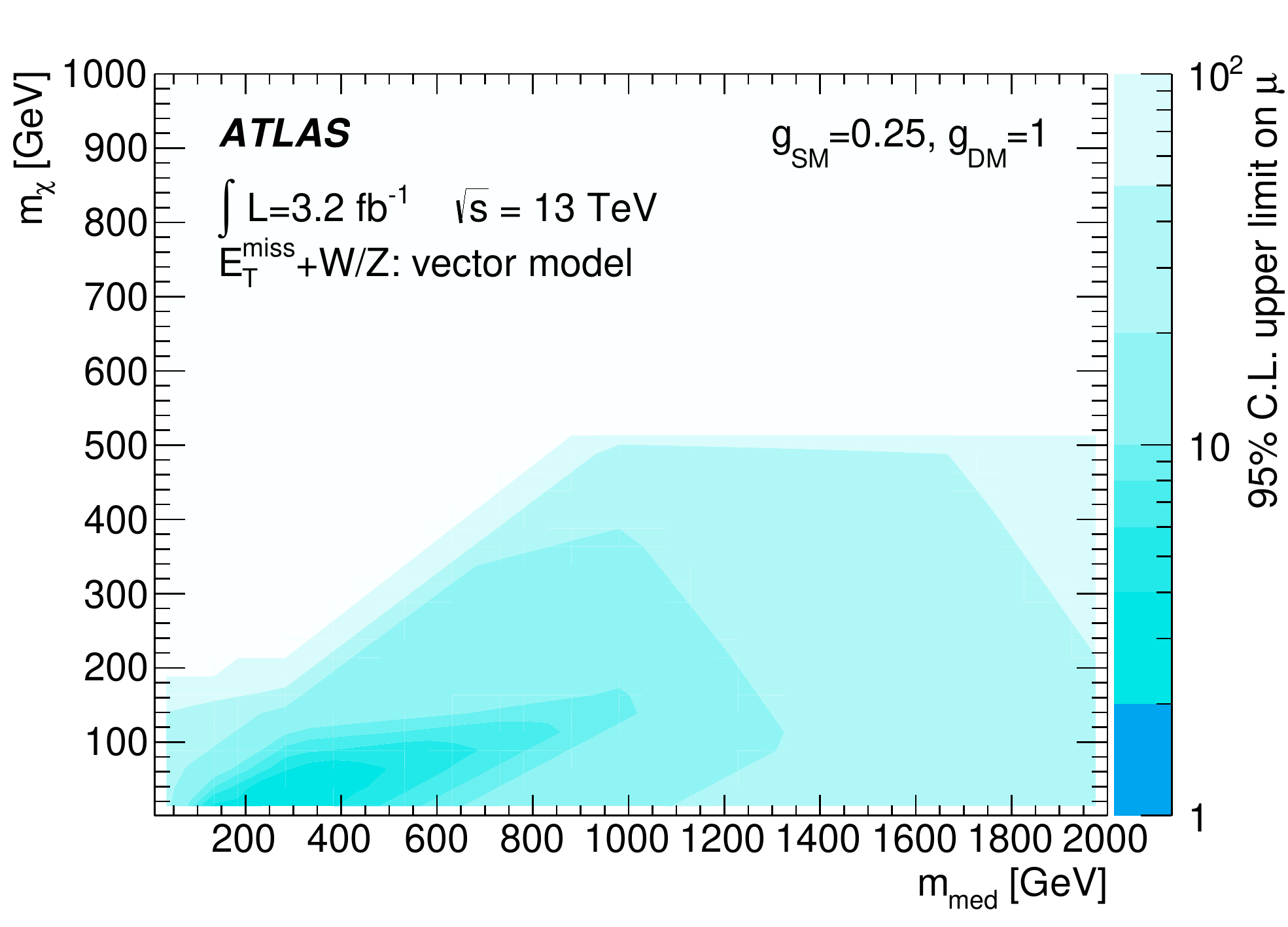} 
}
\caption{
\label{fig:2dLimitPlots} 
Pane (a) shows the limit on the mass scale, $M_{\star}$, 
of the $VV\chi\chi$ EFT model. 
Pane (b) shows the observed limit on the signal strength,
$\mu$, of the vector-mediated simplified 
model in the plane of the dark-matter particle mass, $m_{\chi}$, and the mediator 
mass, $m_{\mathrm{med}}$; white areas indicate an upper limit at $\mu \ge 100$. 
}
\end{center}
\end{figure}

Upper limits at 95\% confidence level (C.L.)  on $\mu$ are calculated using the CL$_s$ method~\cite{Read:2002hq}.
For the $VV\chi\chi$ EFT model, these limits are translated into constraints on the 
mass scale, $M_\star$.   Figure~\ref{fig:2dLimitPlots}(a) shows the limit on the mass scale, $M_{\star}$, in the EFT model, as a function of $m_{\chi}$. Figure~\ref{fig:2dLimitPlots}(b)  shows the limits on the  signal strength, $\mu$, for a vector-mediated simplified model generated  with couplings $g_{\rm SM}=0.25$ and $g_{\rm DM}=1$ in the plane of $m_{\chi}$ and $m_{\textrm{med}}$.

\begin{table}
\begin{center}
\caption{
\label{tab:event_yields}
Predicted and observed number of events in the signal region. 
The yields and uncertainties of the backgrounds are shown after the profile-likelihood fit to the data under the background-only hypothesis. For comparison, the expected yield in the $VV\chi\chi$ EFT model with $M_{\star}=600$ GeV and $m_{\chi}=500$ GeV is $10.1\pm0.4$ events.
}
\vspace{0.5cm}
\begin{tabular}{llr}\hline\hline												
  Process 					 & Events  	\\ \hline 			
  $Z$+jets                   & $\phantom{0}544 \pm 33$        	\\
  $W$+jets                   & $\phantom{0}275 \pm 24$        	\\ 
  $t\bar{t}$ and single-top  & $\phantom{0}211 \pm 19$         	\\
  Diboson                    & $\phantom{00}89 \pm 12$         	\\\hline
  Total Background\ \ \ \ \ \  \  &	$1120   \pm 47   $                             \\
\hline													
		Data  					&	$1121$				\\
\hline\hline													
\end{tabular}
\end{center}
\end{table}

\begin{table}
\begin{center}
\caption{
\label{tab:normalization_factors}
Background normalization factors relative to the initial theoretical prediction, extracted from the profile-likelihood fit under the background-only hypothesis.
}
\vspace{0.5cm}
\begin{tabular}{lr}\hline\hline
Process                   & Normalization Factor \\
\hline
$Z$+jets                  & $1.01 \pm 0.16$ \\
$W$+jets                  & $0.90 \pm 0.16$ \\
$t\bar{t}$                & $0.91 \pm 0.18$ \\
\hline\hline
\end{tabular}
\end{center}
\end{table}

In conclusion, this Letter reports ATLAS limits on dark-matter
production in events with a hadronically decaying $W$  or
$Z$ boson and large missing transverse  momentum. These limits from 3.2 fb$^{-1}$ of 13 TeV $pp$ collisions at the LHC  improve on earlier ATLAS results. No statistically 
significant excess is observed over the Standard Model prediction.

\clearpage


We thank CERN for the very successful operation of the LHC, as well as the
support staff from our institutions without whom ATLAS could not be
operated efficiently.

We acknowledge the support of ANPCyT, Argentina; YerPhI, Armenia; ARC, Australia; BMWFW and FWF, Austria; ANAS, Azerbaijan; SSTC, Belarus; CNPq and FAPESP, Brazil; NSERC, NRC and CFI, Canada; CERN; CONICYT, Chile; CAS, MOST and NSFC, China; COLCIENCIAS, Colombia; MSMT CR, MPO CR and VSC CR, Czech Republic; DNRF and DNSRC, Denmark; IN2P3-CNRS, CEA-DSM/IRFU, France; GNSF, Georgia; BMBF, HGF, and MPG, Germany; GSRT, Greece; RGC, Hong Kong SAR, China; ISF, I-CORE and Benoziyo Center, Israel; INFN, Italy; MEXT and JSPS, Japan; CNRST, Morocco; FOM and NWO, Netherlands; RCN, Norway; MNiSW and NCN, Poland; FCT, Portugal; MNE/IFA, Romania; MES of Russia and NRC KI, Russian Federation; JINR; MESTD, Serbia; MSSR, Slovakia; ARRS and MIZ\v{S}, Slovenia; DST/NRF, South Africa; MINECO, Spain; SRC and Wallenberg Foundation, Sweden; SERI, SNSF and Cantons of Bern and Geneva, Switzerland; MOST, Taiwan; TAEK, Turkey; STFC, United Kingdom; DOE and NSF, United States of America. In addition, individual groups and members have received support from BCKDF, the Canada Council, CANARIE, CRC, Compute Canada, FQRNT, and the Ontario Innovation Trust, Canada; EPLANET, ERC, FP7, Horizon 2020 and Marie Sk{\l}odowska-Curie Actions, European Union; Investissements d'Avenir Labex and Idex, ANR, R{\'e}gion Auvergne and Fondation Partager le Savoir, France; DFG and AvH Foundation, Germany; Herakleitos, Thales and Aristeia programmes co-financed by EU-ESF and the Greek NSRF; BSF, GIF and Minerva, Israel; BRF, Norway; Generalitat de Catalunya, Generalitat Valenciana, Spain; the Royal Society and Leverhulme Trust, United Kingdom.

The crucial computing support from all WLCG partners is acknowledged gratefully, in particular from CERN, the ATLAS Tier-1 facilities at TRIUMF (Canada), NDGF (Denmark, Norway, Sweden), CC-IN2P3 (France), KIT/GridKA (Germany), INFN-CNAF (Italy), NL-T1 (Netherlands), PIC (Spain), ASGC (Taiwan), RAL (UK) and BNL (USA), the Tier-2 facilities worldwide and large non-WLCG resource providers. Major contributors of computing resources are listed in Ref.~\cite{ATL-GEN-PUB-2016-002}.

\printbibliography


\newpage 
\begin{flushleft}
{\Large The ATLAS Collaboration}

\bigskip

M.~Aaboud$^\textrm{\scriptsize 136d}$,
G.~Aad$^\textrm{\scriptsize 87}$,
B.~Abbott$^\textrm{\scriptsize 114}$,
J.~Abdallah$^\textrm{\scriptsize 8}$,
O.~Abdinov$^\textrm{\scriptsize 12}$,
B.~Abeloos$^\textrm{\scriptsize 118}$,
R.~Aben$^\textrm{\scriptsize 108}$,
O.S.~AbouZeid$^\textrm{\scriptsize 138}$,
N.L.~Abraham$^\textrm{\scriptsize 152}$,
H.~Abramowicz$^\textrm{\scriptsize 156}$
% ...
\bigskip
\\
% ...
$^{8}$ Department of Physics, The University of Texas at Arlington, Arlington TX, United States of America\\
% ...
$^{12}$ Institute of Physics, Azerbaijan Academy of Sciences, Baku, Azerbaijan\\
% ...
$^{87}$ CPPM, Aix-Marseille Universit{\'e} and CNRS/IN2P3, Marseille, France\\
% ...
$^{108}$ Nikhef National Institute for Subatomic Physics and University of Amsterdam, Amsterdam, Netherlands\\
% ...
$^{114}$ Homer L. Dodge Department of Physics and Astronomy, University of Oklahoma, Norman OK, United States of America\\
% ...
$^{118}$ LAL, Univ. Paris-Sud, CNRS/IN2P3, Universit{\'e} Paris-Saclay, Orsay, France\\
% ...
$^{136}$ $^{(a)}$ Facult{\'e} des Sciences Ain Chock, R{\'e}seau Universitaire de Physique des Hautes Energies - Universit{\'e} Hassan II, Casablanca; $^{(b)}$ Centre National de l'Energie des Sciences Techniques Nucleaires, Rabat; $^{(c)}$ Facult{\'e} des Sciences Semlalia, Universit{\'e} Cadi Ayyad, LPHEA-Marrakech; $^{(d)}$ Facult{\'e} des Sciences, Universit{\'e} Mohamed Premier and LPTPM, Oujda; $^{(e)}$ Facult{\'e} des sciences, Universit{\'e} Mohammed V, Rabat, Morocco\\
% ...
$^{138}$ Santa Cruz Institute for Particle Physics, University of California Santa Cruz, Santa Cruz CA, United States of America\\
% ...
$^{152}$ Department of Physics and Astronomy, University of Sussex, Brighton, United Kingdom\\
% ...
$^{156}$ Raymond and Beverly Sackler School of Physics and Astronomy, Tel Aviv University, Tel Aviv, Israel\\
% ...
$^{a}$ Also at Department of Physics, King's College London, London, United Kingdom\\
$^{b}$ Also at Institute of Physics, Azerbaijan Academy of Sciences, Baku, Azerbaijan\\
$^{c}$ Also at Novosibirsk State University, Novosibirsk, Russia\\
$^{d}$ Also at TRIUMF, Vancouver BC, Canada\\
$^{e}$ Also at Department of Physics {\&} Astronomy, University of Louisville, Louisville, KY, United States of America\\
% ...
\end{flushleft}


\end{document}